\documentclass[12pt,a4paper]{article}
\usepackage{epsfig}
\usepackage{amsmath}

\def\nostrocostrutto#1\over#2{\mathrel{\mathop{\kern 0pt \rlap 
  {\raise.2ex\hbox{$#1$}}}
  \lower.9ex\hbox{\kern-.190em $#2$}}}
\def\gsim{\nostrocostrutto > \over \sim}   

\catcode`@=11
\newcount\@tempcntc
\def\@citex[#1]#2{\if@filesw\immediate\write\@auxout{\string\citation{#2}}\fi
  \@tempcnta\z@\@tempcntb\m@ne\def\@citea{}\@cite{\@for\@citeb:=#2\do
    {\@ifundefined
       {b@\@citeb}{\@citeo\@tempcntb\m@ne\@citea\def\@citea{,}{\bf ?}\@warning
       {Citation `\@citeb' on page \thepage \space undefined}}%
    {\setbox\z@\hbox{\global\@tempcntc0\csname b@\@citeb\endcsname\relax}%
     \ifnum\@tempcntc=\z@ \@citeo\@tempcntb\m@ne
       \@citea\def\@citea{,}\hbox{\csname b@\@citeb\endcsname}%
     \else
      \advance\@tempcntb\@ne
      \ifnum\@tempcntb=\@tempcntc
      \else\advance\@tempcntb\m@ne\@citeo
      \@tempcnta\@tempcntc\@tempcntb\@tempcntc\fi\fi}}\@citeo}{#1}}
\def\@citeo{\ifnum\@tempcnta>\@tempcntb\else\@citea\def\@citea{,}%
  \ifnum\@tempcnta=\@tempcntb\the\@tempcnta\else
   {\advance\@tempcnta\@ne\ifnum\@tempcnta=\@tempcntb \else \def\@citea{--}\fi
    \advance\@tempcnta\m@ne\the\@tempcnta\@citea\the\@tempcntb}\fi\fi}
\catcode`@=12

\begin{document}

\setcounter{page}{0}
\thispagestyle{empty}

\vfill

\begin{center}
  {\large {\bf

$B$ decays into light scalar particles and glueball\footnote{Work
      supported in part by the Schweizerischer Nationalfonds.}}  }
\vfill
{\bf
    Peter Minkowski } \\
    University of Bern \\
    CH - 3012 Bern, Switzerland
   \vspace*{0.3cm} \\  
   and \vspace*{0.3cm} \\
{\bf
    Wolfgang Ochs } \\
    Max-Planck-Institut f\"ur Physik \\
    Werner-Heisenberg-Institut \\
    D - 80805 Munich, Germany\\  

\vfill
\begin{abstract}
\noindent
The recent observations of $f_0(980)$
in charmless $B$-decays motivate further
studies of scalar particle and glueball production in these processes.
Amplitudes for charmless 2-body $B$ decays involving the members of the scalar
nonet are presented based on the symmetries of the  dominant penguin 
contribution. Different scenarios for the lightest scalar nonet are
investigated in view of the presently available data. We describe the
evidence from $B$-decays for $f_0(1500)$ with a flavour octet like mixing 
and the hints 
towards the members of the $q\bar q$ nonet of lowest mass. 
There is further support
for the hypothesis of a broad $0^{++}$ 
glueball acting as coherent background especially in $B\to K\overline K K$.
The estimated $B$ decay rates into gluonic mesons represent a sizable 
fraction of the theoretically derived decay rate for $b\to sg$.
\end{abstract}
\end{center}  

\vfill



\newpage
\section{Introduction}
A longstanding problem in QCD is the prediction of gluonic boundstates
 (``glueballs'') and the lack of their doubtless identification. 
The lightest glueball is expected in the scalar channel with
$J^{PC}=0^{++}$. In this channel there is a series of 
established resonances but also various broad objects whose existence is not
generally accepted. There is also no 
consensus about the lightest 
scalar $q\bar q$ nonet, neither about its members nor about
the mixing between strange and nonstrange components.
A central issue is the nature of $f_0(980)$ which
has been considered not only as standard $q\bar q$ meson 
but also as $K\overline K$-molecule or as $qq\bar q \bar q$ state.

There are now new experimental results on charmless
$B$ decays into scalar particles which provide 
additional information of high statistics. 
In this paper we discuss
some recent results and their implications on scalar spectroscopy:\\
\vspace*{-0.5cm}
\begin{description}
\item 1. The observation by the BELLE \cite{belle,belle2,belle3,belle4} 
and BaBar collaborations \cite{babar1,babar2}  of charmless 
decays  $B\to K h^+h^-$ with $h=\pi,K$ with a significant peak related to
$f_0(980)$ and some less pronounced signals from other scalars.\\
\vspace*{-0.7cm}
\item 2. In the channel $B\to K \overline K K$ BELLE has also observed a 
broad enhancement (coherent ``background'') 
in the $K\overline K$ mass spectrum in the range 1000-1700
MeV and perhaps beyond with  spin $J=0$ and 
a smaller effect in $\pi\pi$ around 1000 MeV.
Similar results are shown by BaBar 
but there is no quantitative analysis yet.\\
\end{description}
Remarkably, in $B$ decays, as compared  to decays of charmed $D$ or $J/\psi$
mesons
 the role of scalar particles is more pronounced
because of the larger phase space and therefore reduced background 
from crossed channels and also because of an apparent 
 suppression of higher spin ($J=2$) states.
In this paper we will discuss what can be concluded from the above
observations and how further experimental studies can uncover the members
of the light scalar nonet besides $f_0(980)$ and possibly help
to identify the gluonic interaction related to the $0^{++}$ glueball.
Some considerations concerning earlier data have been presented already in
\cite{mozero}. Other pertinent results concern the observation of $f_0(980)$
and gluonic mesons in gluon jets \cite{mozero,moringberg}. Not much theoretical 
work so far has been concerned with $B$-decays into scalar particles.
Some selected $B$-decays into flavoured scalar mesons have been derived 
from a factorization ansatz for the effective weak Hamiltonian including 
power corrections ${\cal O}(\Lambda/m_b)$ \cite{chernyak}.

The interest in charmless  $B$-decays with strangeness has been stimulated
through the observation by
CLEO \cite{cleo0,cleo1} of large inclusive and exclusive 
decay rates $B\to \eta'X$ and $B\to \eta'K$, which have been
confirmed by more recent measurements \cite{cleo2,belle1,babar}. These
processes have been related to the decay  $b\to sg$ of the $b$-quark 
which could be a source of mesons with large gluon affinity
\cite{soni,fritzsch,hou,dgr}. In consequence, besides $\eta'$ 
also other gluonic states, in particular also scalar mesons or glueballs  
could be produced in a similar way. 

The total rate $b\to sg$  has been calculated  perturbatively
in leading \cite{ciuchini} and next-to-leading order \cite{greub}
\begin{equation}
{\cal B} (b\to sg) = 
\begin{cases}
(2-5)\times 10^{-3} & \text{in LO  (for $\mu=m_b\ldots m_b/2$)}\\
(5\pm 1)\times 10^{-3} &  \text{in NLO}
\end{cases}
\label{btosg}
\end{equation} 
The energetic massless gluon in this process could turn
entirely into gluonic mesons by a nonperturbative transition
after colour neutralization by a second gluon.
Alternatively, colour neutralization through $q\bar q$ pairs is
possible
as well. This is to be distinguished from the short distance process $b\to   
s\bar q q$ with virtual intermediate gluon
which has to be added to the
CKM-suppressed decays $b\to q_1\bar q_2 q_2$. These quark processes
with $s$
have been calculated and amount to branching fractions of 
$\sim 2\times 10^{-3}$ each
\cite{altarelli,nierste,greub}.
The question then arises which 
hadronic final states correspond to the decay  $b\to sg$.

We recall that a large gluonic penguin component has been suggested to
play also an important if
not dominant role in the explanation of the $\Delta
I=\frac{1}{2}$ rule in $K$ decay \cite{pm}.

Next we outline the status of the phenomenological discussion concerning the
light scalar spectroscopy. The Particle Data Group \cite{pdg} lists below
1800 MeV the following isoscalar particles
\begin{equation}
f_0(600) (\text{or} \sigma),\ f_0(980),\ f_0(1370),\ f_0(1500),\ f_0(1710)
\end{equation}
furthermore the isovectors $a_0(980),\ a_0(1450)$ and the strange
$K^*_0(1430)$, possibly there are $\kappa(850)$ and $K^*_0(1950)$.
The broad states $\sigma,\kappa$ are still controversial, also not much reliable
information is available about $f_0(1370)$.
The  states listed above 
should be related to a scalar $q\bar q$ nonet and, possibly, a glueball.

Quantitative results on glueballs are derived today from the QCD lattice calculations or
QCD sum rules, both agree that the lightest glueball has quantum numbers
 $J^{PC}=0^{++}$.

Lattice calculations in
quenched approximation
\cite{bali,svw,Morning,lt} (without light sea quark-antiquark pairs)
suggest
the lightest glueball to have a mass in the range 1400-1800 MeV
\cite{balirev}. Results from unquenched calculations still suffer from 
systematic effects, the large quark masses of the order of the strange
quark mass and large lattice spacings. Typically, present results
on the glueball mass are about 20\% lower
than the quenched results \cite{ht,hnm}.

Results on glueballs have also been obtained from QCD sum rules.
Recent calculations \cite{narison} for the  $0^{++}$ glueball
yield a mass consistent with the quenched lattice result
but in addition require a gluonic state near 1000 MeV. 
Similar results with a low glueball mass around 1000 MeV are obtained
also in other calculations \cite{steele}. On the other hand, it has been
argued \cite{forkel} that the sum rules
can also be saturated by a single glueball
state with mass $1250\pm 200$ MeV. 

In conclusion, there is agreement in the QCD based calculations on the
existence of a $0^{++}$ glueball but the mass and width of the lightest
state
is not yet certain 
and phenomenological searches should allow a mass range of about
1000-1800 MeV.  

In the interpretation of the phenomenological results one can identify two
major different directions of thought; they differ in the role of $f_0(980)$
which belongs either to a nonet with particles of higher mass or with
particles  of lower mass.

\noindent {\it Route A: Scalar nonet with $f_0(980)$ and heavier particles}\\
In a previous study \cite{mo} 
we have performed 
a detailed phenomenological analysis of production
and decay of low mass scalar mesons, which led us to identify 
the lightest $q\bar q$ scalar nonet with the states
\begin{equation}
f_0(980),\ a_0(980),\ K^*_0(1430),\ f_0(1500) \label{moscnonet}
\end{equation}
 with large flavour
mixing, just as in the pseudoscalar nonet, i.e. with flavour components 
($u\bar u, d\bar d, s\bar s$) approximately given by
\begin{equation}
\eta',\ f_0(980)\ \leftrightarrow \ (1,1,2)/\sqrt{6},\qquad \eta,\ f_0(1500)
 \leftrightarrow \ (1,1,-1)/\sqrt{3} \label{mononet}
\end{equation}
close to the flavour singlet or octet respectively and 
with the parity partners $\eta'$ and
$f_0(980)$.

This scalar nonet separates to a good approximation into the singlet
$f_0(980)$ and the octet $a_0(980),\ K^*_0(1430),\ f_0(1500)$. The octet,
within this approximation,
fulfills the Gell Mann-Okubo mass formula 
and is also consistent with a chiral model with general QCD interaction of
$\Sigma$ fields.
In elastic  $\pi\pi$ scattering and in other channels the cross section
shows  a broad
``background'' which extends from below 1000 MeV up to about 1600 MeV or higher,
with dips from negative interference with narrow states $f_0(980)$ and
$f_0(1500)$ (``red dragon''). The PDG has been listing here the two states
left over so far $f_0(600)$ 
and $f_0(1370)$. We have interpreted this ``background''
as signals from a single broad object $f_0(1000)$ centered around 1000~MeV with 
a large width of   
500-1000 MeV (according to the $T$ matrix pole parameters \cite{hyams,mp}).
This ``left over state'' was taken as $0^{++}$ glueball as its production
characteristics followed largely the expectations for a gluonic meson 
(except for $J/\psi\to \gamma \pi\pi$).  
Subsequent studies of various decay rates and relative amplitude phases 
involving $f_0(980)$ and $f_0(1500)$ 
have confirmed this view \cite{momont1,momont2},
sometimes better experimental data would be desirable. The
light scalar objects $\sigma,\kappa$ are not considered as individual physical
particles to be classified into a nonet. The heavier particles $a_0(1450),\ 
f_0(1710)$ could be members of a second nonet, possibly with $K^*(1950)$ and
$f_0(2020)$.

There are other approaches which agree on $f_0(980)$ as being the lightest
particle in the nonet. 
A scheme similar to ours above 
for the $q\bar q$ nonet with $f_0(980)$ and $f_0(1500)$ and mixing like
(\ref{mononet})
has been proposed  \cite{klempt} based on a quark model with instanton
interactions. It prefers though  $a_0(1450)$ over $a_0(980)$ as nonet
partner of $f_0(980)$, but there is no prediction on a glueball.
The broad enhancements in $\pi\pi$ and also $4\pi$ spectra have been 
interpreted as a consequence of $\rho$-exchange in two-body 
scattering amplitudes \cite{klemptrev}.

A broad glueball (width about 2 GeV) is found in the mass range
1200-1600 MeV from overlapping $f_0$ states in a $K$ matrix analysis 
of a variety of reactions \cite{anis}. The glueball nature of this state
has been inferred from its ``flavour blind couplings''. Two $q\bar q$
multiplets emerge in the mass range below 1900 MeV including $f_0(1300)$
and again $f_0(980)$ as the lightest scalar which now  behaves like a 
flavour octet.

Using results from QCD sum rules a scheme for light scalars 
has been proposed \cite{narison} where a broad
isoscalar ``$\sigma$'' around 1000 MeV and the narrow $f_0(980)$ are 
both mixed in 
equal parts from the broad glueball and a light $q\bar q$ scalar. 

Despite various differences in detail, especially the role of the glueball, 
all these schemes have $f_0(980)$ as the lightest member of the nonet.

\noindent {\it Route B: Scalar nonet with $f_0(980)$ and lighter
particles}\\
In a second line of thought there is one $q\bar q$ nonet at heavier mass
and, in general, a second nonet $q\bar q$ or $qq\bar q\bar q$ at lower mass
including $f_0(980)$. The heavier multiplet includes $K^*_0(1430)$,
the isovector candidate in the same mass region is $a_0(1450)$. In the
isoscalar channel one observes $f_0(1370),\ f_0(1500)$ and $f_0(1710)$.
The scalar glueball is assumed with mass around 1600 MeV as suggested by
quenched lattice calculations.
Then the glueball and two isoscalar members of the nonet can mix and
generate the three observed $f_0$ states above. Several such mixing schemes
have been proposed giving either a larger gluonic component to $f_0(1500)$
\cite{amsler} or to $f_0(1710)$ \cite{lee}. 

The lower mass scalars are now left over. 
An interesting possibility is
the existence of a light nonet including 
\begin{equation}
f_0(980),\  a_0(980),\ \kappa(850),\ \sigma(600) \label{scalarB}
\end{equation}
(alternatively also $K^*(1430)$).
This nonet could be of conventional $q\bar q$ type
\cite{morgan,toern,scadron,ishida} or built from $qq\bar q\bar q$
\cite{jaffe,achasov,schechter}. 
Mesons of this light nonet can also be related to poles in meson
meson scattering amplitudes constructed using chiral symmetry and unitarity
\cite{oller}. Accepting the existence of $\sigma$ and $\kappa$ particles one
could have two nonets below 1800 MeV (for reviews, see also
\cite{closetoern,tuan,amslertoe}). 

Finally, we should comment on the problem of the $\sigma,\ \kappa$ poles
whose interpretation as particles remains controversial \cite{penn,wo}.
The evidence
has been studied in detail in $\pi\pi$ scattering  
within a class of parametrizations which 
respect chiral
symmetry  and unitarity \cite{cgl}. The
small scattering length and the  rapid increase of the isoscalar $S$ wave phase
shifts with energy required by unitarity yields a pole in the amplitude 
related to  $\sigma$.
Whether the pole in this parametrization represents a physical
particle is not immediately obvious. 

In particular, it has been pointed out \cite{meissner} that final state
$\pi\pi$ interactions calculated within the non-linear sigma model
up to two loops "mock up" such a particle: the ``$\sigma$'' particle
is nothing but a convenient parametrization of the correlated two pion
exchange.
Here we wish to argue that
-- if these objects are really physical particles --    
the decay amplitudes should respect the associated
flavour symmetry relations which we discuss below.

The aim of our paper is the identification of the lightest scalar nonet
and possibly the light scalar glueball. The first step
in the next Section is a classification of amplitudes in charmless B decays. 
We describe 
an approximation for decays into two pseudoscalar (PP) or into
a pseudoscalar and a vector particle (PV) employing U(3) flavour symmetry
and compare with the recent experimental results;
gluonic transitions play an important role. In Section 3 we turn to the 
discussion of scalar particles (S) in final states (PS) and (VS) which is the 
main interest of the paper and we present the corresponding decay amplitudes.
For illustration we compare our formulae with the first 
experimental data for two different scenarios of the scalar sector and 
present further predictions. One possible 
interpretation of the data involves a broad scalar glueball; in Section 4
we estimate the total gluonic contributions to B decays 
and compare with perturbative calculations. Conclusions follow in Section 5.
The detailed study of charmless $B$ decays should help distinguishing the
various possibilities outlined above. 

\section{Charmless $B$ decays with $K$ and $K^*(890)$}
Our aim is the description of charmless $B$ decays into scalar particles (S) 
together with particles from pseudoscalar (P) and vector (V) 
multiplets, $B\to PS,\ B\to VS$, including especially the recently observed
scalar $f_0(980)$. In view of the yet incomplete understanding of the scalar
sector and the scarce data we are interested in 
the leading approximation which describes the main effects.
We begin by reconsidering the well studied decays $B\to PP, B\to VP$,
especially $B\to K\eta',\ K^*\eta'$, 
together with the other final states related 
by $U(3)$ symmetry. Subsequently we extend these considerations 
towards the inclusion of scalar particles.

\begin{figure}[t!]
\begin{center}
\vspace{-1.5cm}
\includegraphics[angle=+90,width=15cm]{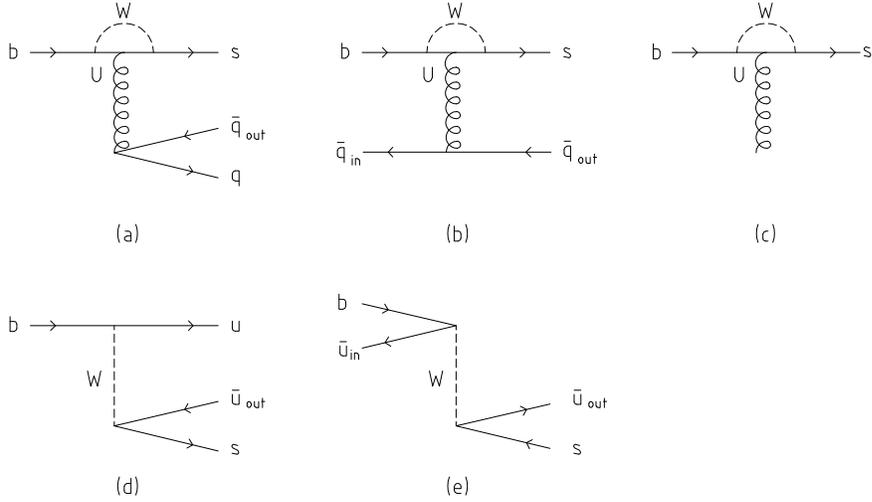}
\end{center}
\vspace{-2.5cm}
\caption{Short distance processes for charmless $b$ decays
(a) QCD penguin diagram with  $q\bar q$ pair production 
through intermediate
gluon ($U$ represents $u,c,t$ quarks) 
and (b) with spectator annihilation; (c) penguin diagram with hard
gluon; (d) electroweak (CKM suppressed) tree diagram for $q\bar q$ pair
production and (e) spectator annihilation. In the present analysis we keep
only (a) and (c).
}
\label{shortdist}
\end{figure}    


\subsection{Approximation for two-body decays}
The large branching fraction $B\to K\eta'$
confirms the special role of $\eta'$ in these
decays and it has been related \cite{soni,fritzsch,hou,dgr} to the gluon
affinity of $\eta'$, especially 
through the QCD axial anomaly which affects only the flavour singlet
component. In the factorization approach many 2-body decays can be
reproduced well but not $B\to K\eta'$ \cite{alikramer,ag}. Also, it
appears difficult to explain the $K\eta'$ rate entirely by quark final
states and the QCD anomaly within a perturbative framework \cite{acgk},
a factor 2 remains unexplained. An improvement is possible by inclusion of
QCD radiative corrections \cite{bn} but with considerable uncertainties.

Alternatively, one may introduce a phenomenological flavour singlet
amplitude which allows also for non-perturbative effects \cite{dgr}.
This amplitude is added to the dominant 
penguin amplitudes, 
 the small 
 tree amplitudes and electroweak penguins. Different decays are related by
flavour $U(3)$ symmetry.
 Recent applications \cite{cr,cr1} of this scheme to 2-body $B$ decays  
with strange and nonstrange pseudoscalar and vector  
particles yield a good overall
agreement with the data in terms of a few phenomenological input amplitudes. 

Here we discuss
first the 2-body $B$ decays with $K$ and $K^*$ in this way \cite{cr,cr1}
to understand the pattern of the observed rates and then extend the analysis
to the scalar sector. 
For this purpose we restrict ourselves to a simple approximation. 
In the description of the short distance interaction we keep only
the dominant QCD penguin amplitudes $T_q$ for $b\to q\bar q s$,
 $q=u,d,s$  with intermediate virtual gluon and 
$T_u= T_d=T_s$ as well as the penguin amplitude with  hard 
gluon radiation (see Fig.
\ref{shortdist}a,c) while we neglect the electroweak tree diagrams which are
CKM suppressed in $|V_{us}|=0.22$ (Fig.~\ref{shortdist}d)
as well as the spectator
annihilation processes of any kind (Fig.~\ref{shortdist}b,e) because of the
large $b$-quark mass. 

Concerning the suppression of the tree diagrams we refer to the calculations
\cite{nierste,greub} of
decay rates at the quark level including penguin and
tree amplitudes. The
decays of interest to us 
are found with relative fractions \cite{nierste} 22\% ($b\to u\bar u
s$), 18\%
($b\to d\bar d s$) and 15\% ($ b\to  s\bar s s$)
 of all charmless hadronic 
$B$ decays (a total of $(55\pm 5\%$), the remaining $\sim$
45\% correspond to $b\to sg$). So 
the nonleading weak decay amplitudes which we neglect here 
modify the leading result for these rates 
from penguin amplitudes by about $\pm20\%$; they become essential, if CP
violating effects are investigated.

\begin{figure}[t!]
\begin{center}
\vspace{-1.5cm}
\includegraphics[angle=+90,width=15cm]{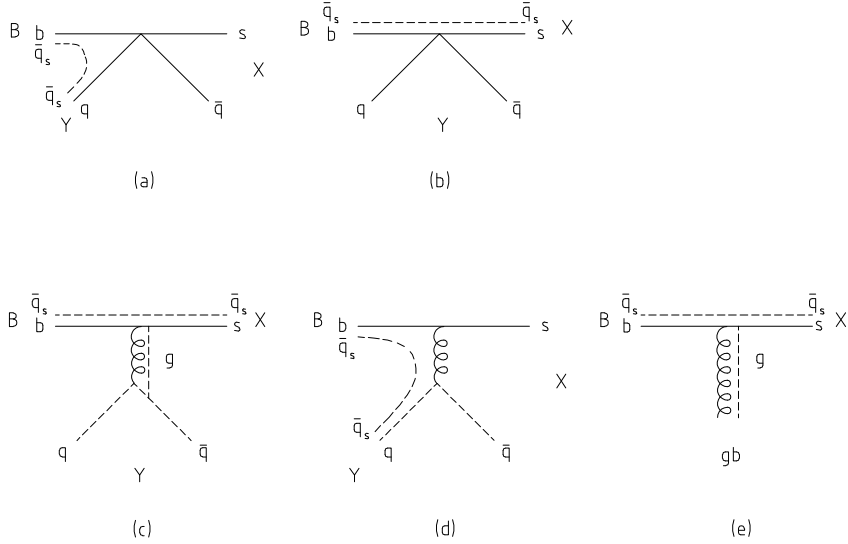}
\end{center}  
\vspace{-2.0cm}
\caption{
Diagrams for charmless hadronic two-body decays $B\to xy$ 
(likewise $B\to yx$) 
into hadrons $x, y$ from flavour multiplets $X,Y$ 
related to QCD penguin
amplitudes of Figs. 1a-c:  with  $b\to s \bar q q$ followed by (a) colour favoured 
and (b) colour suppressed hadron formation;
 with $b\to sg$  followed by (c) gluonic colour neutralization and (d)
 $q\bar q$ colour neutralization or (e) glueball production. 
The dashed lines 
indicate quarks and gluons with typically soft interactions.
}
\label{hadrons}
\end{figure}
Next we discuss the hadronic two-body decays $B\to xy$ 
with particles $x,y$ belonging to
 $U(3)$ multiplets $X$ and $Y$ which are related to the penguin diagrams
Fig.~\ref{shortdist}a,c kept in the present approximation.
The $b\to sq\bar q$ amplitudes in Fig.~\ref{shortdist}a lead to
hadrons either by connection of the spectator quark $(q_s)$ in the $B$ meson 
 with the produced quark $q$ or with the $s$ quark as shown in Figs.
\ref{hadrons}a,b. The second process is colour suppressed as the 
$q\bar q$ pair produced in a colour octet state cannot recombine directly into
a hadron; we will neglect this contribution in the following.
The hadronic penguin amplitudes $p_{xy}^q$ for the decay 
$B\to x(s\bar q)y(q\bar q_s)$ into particles from multiplets $X$ and $Y$ 
in Fig.~\ref{hadrons}a  
are then assumed proportional within the given multiplet 
to the amplitudes
$T_q$ in Fig.~\ref{shortdist}a 
\begin{equation}
p_{xy}^q=A^x_{s\bar q}A^y_{q\bar q_s} p_{XY},\quad p_{XY}=h_{XY}T,\quad T\equiv T_u=T_d=T_s
\label{pXY}
\end{equation}
where $A^x_{q\bar q'}$ denotes the flavour coupling  $x\to q\bar q'$ 
 and  $h_{XY}$ a hadronization constant. Likewise there are
amplitudes $p_{yx}^q$ with  $x$ and $y$ exchanged.

\begin{figure}[t]
\begin{center}
\vspace{-3.0cm}
\includegraphics[angle=+90,width=15cm]{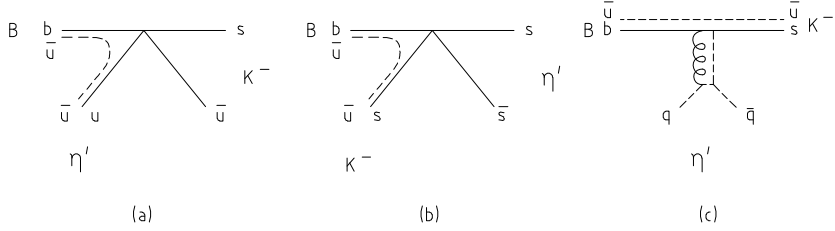}
\end{center}  
\vspace{-5.0cm}
\caption{
Two-body decay $B^-\to K^-\eta'$ with three amplitudes as in Fig.
\ref{hadrons}a,c:
 (a) amplitude $p^u_{K^-\eta'}$ with  $s\to K^-$, (b) exchange
amplitude $p^s_{\eta'K^-}$ with
 $s\to \eta'$ and (c) amplitude $s_{K^-\eta'}$ for 
gluonic production of $\eta'$.}
\label{Ketaprime}
\end{figure}

In addition, there is the contribution from the $b\to sg$ penguin
in Fig.~\ref{shortdist}c which leads to hadronic final states 
either by non-perturbative soft 
colour octet ($g,gg,\ldots)$ or by colour triplet ($q\bar q$)
neutralization of the hard gluon as in Fig.
\ref{hadrons}c,d.\footnote{For a further discussion of these 
two mechanisms and observable consequences, see, for example, Ref. 
\protect\cite{moringberg}.}
The amplitude in Fig. \ref{hadrons}c  contributes to the
production of a meson $y$ with flavour singlet $q\bar q$ component
(such as the $\eta'$)
which we write as 
\begin{equation}
s_{xy}=A^x_{s\bar q_s}(A^y_{u\bar u}+A^y_{d\bar d}+A^y_{s\bar s}) %
s_{XY},\quad \text{with}
  \quad  s_{XY}=\gamma_{XY} p_{XY}.
\label{sXY}
\end{equation}
We take into account this
contribution only for the production of 
scalar and pseudoscalar particles and neglect it for vector particles
($\omega,\phi$).
This can be justified by the smaller perturbative transition amplitude 
with exchange of at least three gluons in the second case as compared
to two gluons in the first one; this difference is also considered responsible for 
the large flavour mixing in the pseudoscalar and the small mixing in the 
vector meson nonet \cite{fm}. The smallness of this flavour singlet
contribution has also been found in the phenomenological analysis \cite{cr1}.
Furthermore, glueball production is possible
through this process (Fig.~\ref{hadrons}e). The diagram in Fig.~\ref{hadrons}d
 has the same quark structure as the one in Fig.~\ref{hadrons}a and therefore 
it is not kept independently. 

If the two particles belong to two different multiplets then $p_{yx}^q$
with both decay particles interchanged, i.e. with  $s\to y,\ q_s\to x$,
is written as $\beta_{XY}' p_{xy}^q$.
In case of $B$-decays into an isoscalar meson 
with mixed strange and non-strange quark
components ($\eta,\eta',\ldots$) both amplitudes contribute and
interfere which yields the
full amplitude $p_{xy}^q+ \beta_{XY}p_{xy}^{q'}$ 
where the second term with
$\beta_{XY}=(-1)^L\beta_{XY}'$ 
refers to the 2-particle state with reflected momenta
($\vec p \to -\vec p$) for orbital angular momentum $L$
\cite{lipkin}. In particular, for 
$B\to VP$ decays there is a relative $(-)$ sign for the interchange amplitude.

The $B$ branching ratios are then computed from the superposition
of up to three amplitudes depending on the quark structure of $x$ and $y$:
the amplitude for decay $B\to x(s\bar q)y(q\bar q_s)$, the exchange amplitude
for decay $B\to y(s\bar q')x(q'\bar q_s)$ 
and the gluonic amplitude for decay into
an isoscalar meson $B\to x(s\bar q_s)\ y(\text{isoscalar}\ q\bar q$).
\begin{equation}
\begin{split}
{\cal B}(B& \to xy) \\
 =&|p_{xy}^{q}+p_{yx}^{q'}+s_{xy}|^2\\
 =&|h_{XY}|^2|A^x_{s\bar q}A^y_{q\bar q_s}T_q+
      A^y_{s\bar q'}A^x_{q'\bar q_s}\beta_{XY}T_{q'}+
A^x_{s\bar q_s} \sum_{q''}A^y_{q''\bar q''}\gamma_{XY}T|^2.
\end{split}
\label{braratio}
\end{equation}
with produced quarks $q,q',q''$.
The  decay $B^-\to K^-\eta'$ 
where all three amplitudes contribute is shown in Fig.
\ref{Ketaprime}.
For the nonets of pseudoscalar, vector and scalar mesons
considered here we list the relevant parameters and some notations 
in Table~\ref{tab:notation}.

\begin{table}[t!]
\caption{Parameters for decays $B\to xy$ ($s\to x,\ q_s\to y$) 
for pseudoscalar (P), vector (V) and scalar (S) particles and some special
choices in the present analysis}
$
\begin{array}{lllll}
\hline\hline
XY \quad & \text{penguin} & 
      & \text{flavour singlet} & \text{special choices} \\ 
\hline
PP & p_{PP} &         & s_{PP}=\gamma_{PP}p_{PP}& \\
VP & p_{VP} & p_{PV}=\beta_{VP} p_{VP}  & s_{VP}=\gamma_{VP} p_{VP} & 
      \beta_{VP}=-1,\ \gamma_{VP}=\gamma_{PP}\\
PS & p_{PS} &  p_{SP}=\beta_{PS} p_{PS}& s_{PS}=\gamma_{PS} p_{PS} & 
   \beta_{PS}=\pm1 \\
 & & & s_{SP}=\gamma_{SP} p_{PS}& \\ 
VS & p_{VS} &  p_{SV}=\beta_{VS} p_{VS}& s_{VS}=\gamma_{VS} p_{VS} & 
  \beta_{VS}=-\beta_{PS},\ \gamma_{VS}=\gamma_{PS} \\
\hline\hline
\end{array} 
$
\label{tab:notation}
\end{table}

\vspace{0.5cm}
\subsection{Decays into pseudoscalar and vector particles}

We apply this simple approximation based on the dominance of penguin
diagrams first to the well studied $B\to PP$ and $B\to VP$ decays. Non-leading
diagrams are considerd in \cite{cr,cr1} which are important for the study of
CP violation effects (absent in our approximation). The relative magnitudes
of the tree amplitudes are found there of the order of $\sim$20\%.

Without loss of generality we can define $p_{XY}$
as real in our approximation, whereas  $\beta_{XY}$ and $\gamma_{XY}$ 
could be complex in general. However, we assume in the present application
also  $\beta_{XY}$ and $ \gamma_{XY}$ to be real, to begin with. Furthermore,
we assume equal  recombination probability for the exchanged particles, 
i.e. $|\beta_{XY}|=1$ or 
$\beta_{XY}=\pm(-1)^L$.
These simplifications could be relaxed if required by the data.
The quark mixing parameters in the pseudoscalar sector are taken from 
(\ref{mononet}), i.e. $\eta=(u\bar u + d\bar d - s\bar s)/\sqrt{3}$ and
$\eta'=(u\bar u + d\bar d + 2 s\bar s)/\sqrt{6}$.

\begin{table}[p]
\caption{
Branching ratios for $B^+$ and $B^0$ decays into pseudoscalar (P) and
vector (V) particles (cols. 4,5) and amplitudes in Eq. (\ref{braratio}) 
(col. 2), 
$\gamma_{PP},\gamma_{VP}$ and $\beta_{VP}$ 
  for gluonic and interchange processes; 
 col. 3: $T_q,\alpha\equiv p_{VP}/p_{PP}$ set to 1;
col.~4: $\alpha=0.661,\ 
\gamma_{PP}=\gamma_{VP}=0.439$, $\beta_{VP}=-1$, $|p_{PP}|^2=20.6\times
10^{-6}$.}
\vspace*{0.1cm}
$
\begin{array}{llclc}
\hline\hline
B\to PP & \text{amplitudes} & T_q=1 &  {\cal B}_{\text{th}} [10^{-6}]
     &   {\cal B}_{\text{exp}} [10^{-6}]\\
\hline
K^0\pi^+ & T_d & 1 &   \text{input}\ p_{PP} & 20.6\pm 1.3
\\
K^+\pi^0 & \frac{1}{\sqrt{2}} T_u & \frac{1}{\sqrt{2}} &   10.3 & 
     12.8\pm 1.1\\
K^+\eta & \frac{1}{\sqrt{3}} (T_u-T_s+\gamma_{PP} T) & 
    \frac{1}{\sqrt{3}}\gamma_{PP} 
     &  1.3 & 3.1\pm0.7
\\
K^+\eta' & \frac{1}{\sqrt{6}} (T_u+2T_s+4\gamma_{PP} T) &
\frac{1}{\sqrt{6}}(3+4\gamma_{PP})      &   \text{input}\ \gamma_{PP}
&77.6\pm4.6
\\
\hline
K^+\pi^- & T_u & 1 & 19.0 & 18.2\pm0.8
\\
K^0\pi^0 & -\frac{1}{\sqrt{2}} T_d & -\frac{1}{\sqrt{2}} & 9.5 & 
11.2\pm1.4
\\
K^0\eta & \frac{1}{\sqrt{3}} (T_d-T_s+\gamma_{PP} T) & 
       \frac{1}{\sqrt{3}}\gamma_{PP} 
    &  1.2 & <4.6
\\
K^0\eta' & \frac{1}{\sqrt{6}} (T_d+2T_s+4\gamma_{PP} T) &
\frac{1}{\sqrt{6}}(3+4\gamma_{PP})      &  71.5 &  60.6\pm7.0
\\
\hline\hline
B\to VP & &\alpha=1,\ \beta_{VP}=-1 & &\\
\hline
K^{*0}\pi^+ & \alpha T_d &  1  & \text{input}\ \alpha & 9.0\pm1.4
\\
K^{*+}\pi^0 &  \frac{\alpha}{\sqrt{2}} T_u & \frac{1}{\sqrt{2}} & 
   4.5 & <31 
\\
K^{*+}\eta & \frac{\alpha}{\sqrt{3}} (T_u-\beta_{VP} T_s+\gamma_{VP} T) &
  \frac{1}{\sqrt{3}}(2+\gamma_{VP})
      &  17.8 &  25.9\pm3.4
\\
K^{*+}\eta' & \frac{\alpha}{\sqrt{6}} (T_u+2\beta_{VP} T_s+4\gamma_{VP} T) &
\frac{1}{\sqrt{6}}(-1+4\gamma_{VP}) &  0.9 & <12
\\
\rho^+ K^0 & \alpha\beta_{VP} T_d & -1 &  9.0 & <48 \\
\rho^0 K^+ &  \frac{\alpha}{\sqrt{2}}\beta_{VP}T_u & -\frac{1}{\sqrt{2}}&
     4.5 &  4.1\pm0.8 \\
\omega K^+&  \frac{\alpha}{\sqrt{2}}\beta_{VP}T_u & -\frac{1}{\sqrt{2}}&
       4.5 & 5.4\pm0.8 \\
\phi K^+ & \alpha T_s & 1   & 9.0& 9.0\pm0.9\\
\hline
K^{*+}\pi^- & \alpha T_u & 1 &   8.3 & 15.3\pm3.8
\\
K^{*0}\pi^0 & -\frac{\alpha}{\sqrt{2}} T_d & -\frac{1}{\sqrt{2}} & 4.2 & 
   0.4\pm1.8
\\
K^{*0}\eta & \frac{\alpha}{\sqrt{3}} (T_d-\beta_{VP} T_s+\gamma_{VP} T) & 
    \frac{1}{\sqrt{3}}(2+\gamma_{VP})   & 16.4 & 17.8\pm2.0
\\
K^{*0}\eta' & \frac{\alpha}{\sqrt{6}} (T_d+2\beta_{VP} T_s+4\gamma_{VP} T) &
\frac{1}{\sqrt{6}}(-1+4\gamma_{VP})     &   0.8&  <6.4
\\
\rho^- K^+& \alpha\beta_{VP} T_u & -1 & 8.3  & 9.0\pm2.3\\
\rho^0 K^0&  -\frac{\alpha}{\sqrt{2}}\beta_{VP}T_d & \frac{1}{\sqrt{2}}&
       4.2 & <12.4 \\
\omega K^0&  \frac{\alpha}{\sqrt{2}}\beta_{VP} T_d & -\frac{1}{\sqrt{2}}&
       4.2 & 5.2\pm1.1 \\
\phi K^0 & \alpha T_s & 1 &  8.3 & 7.8\pm 1.1\\
\hline\hline
\end{array}
$
\label{tab:pseudosc}
\end{table}
Results for $B\to PP$ and $B\to VP$ decays are given in
 Table \ref{tab:pseudosc}, 
the amplitudes entering Eq. (\ref{braratio}) in col. 2 and explicitly in
col. 3 for $T_q=1$  in units of $p_{PP}$ and $p_{VP}$
with parameters $\gamma_{PP}$ and $\gamma_{VP}$ and $\beta_{VP}=-1$.
Repeating the calculation with $\beta_{VP}=+1$ would exchange the roles of
$K^{*+}\eta$ and $K^{*+}\eta'$ and can therefore be 
excluded by comparing with the data.
The $B^+$ decay rates are obtained by adjusting in each sector the
normalization of $|p_{XY}|^2$ and by multiplying the amplitude squared in
col. 3 with $|p_{XY}|^2$.
The predictions for $B^0$ follow by
multiplying $|p_{XY}|^2$ with the ratio $\tau_{B^0}/\tau_{B^+}=0.921$
 \cite{pdg}. 
%
We compare with branching ratios updated recently for $B\to PP$ \cite{bn}
and $B\to VP$ \cite{cr1} in col. 5. 

In a first approximation we adjust only the two penguin amplitudes
$p_{PP}$ and $p_{VP}\equiv\alpha p_{PP}$ from two rates
($K^0\pi^+,K^{*0}\pi^+$) and set
$\gamma_{XY}=0$.
Then the overall pattern of the data 
is reproduced, except for the rates $K\eta'$ and $K^*\eta$ which are 
significantly too large by a
factor of up to $\sim 3$ ($K^+\eta'$: 30.9, $K^0\eta'$: 28.5,
$K^{*+}\eta$: 12.0, $K^{*0}\eta$: 11.1). This conclusion follows here from 
flavour symmetry.
The neglect of nonleading short distance terms, estimated to 
amount about 20\%
cannot account for the discrepency, which is then attributed to the 
additional flavour singlet
amplitudes $\gamma_{PP} p_{PP}$ and  $\gamma_{VP} p_{VP}$. 

Predictions for real $\gamma_P=0.439$, determined from $K^+\eta'$ 
are shown in col. 4, in the VP sector
we chose for simplicity $\gamma_{VP}=\gamma_{PP}$ (see also the similar
results in \cite{cr}).
At this level of approximation, with expected accuracy of $\sim20\%$, 
there are no major discrepencies encountered.\footnote{There 
are two predictions
with a $\sim$2.5$\sigma$ deviation from data ($K^+\eta$ and $K^{*+}\eta$); 
a better result could be obtained by
fitting the three parameters to all branching ratios instead of three only. 
Alternatively, one may relax the
condition for $\gamma_{PP}$ to be real or the condition
$\gamma_{VP}=\gamma_{PP}$.} 

Whereas we do not intend to go beyond the present approximations, we may
estimate the effect of chosing $\gamma_{PP}$ complex. Then we find 
from $K\eta$ and $K\eta'$
rates (Table \ref{tab:pseudosc})
$\gamma_{PP}\sim 0.67\exp(i\varphi_{PP}),\ \varphi_{PP}\sim \pm 67^\circ$.   
Using either real or complex $\gamma_{PP}$ we obtain  an estimate of the 
gluonic part of the $K\eta',\ K\eta$ production rate 
\begin{equation}
{\cal B}(B^+\to K^+ \eta',\ K^+\eta)|_{\text{gluonic}} = 
3\ |\gamma_{PP}\ p_{PP}|^2 
\sim (12\ldots 28)\times 10^{-6}.
\label{etaprgluon}
\end{equation}

We conclude that the main effects are
reproduced in each sector by two parameters, the penguin amplitude $p_{XY}$
and the flavour singlet parameter $\gamma_{XY}$,
furthermore we have chosen $\beta_{VP}=-1$.

Of particular importance for our further discussion are the large and small 
rates for $K\eta'$ and $K\eta$ respectively and the abundancies 
of $K^*\eta'$ and $K^*\eta$ in reversed order
which the model explains after choosing $\beta_{VP}=-1$, 
so far consistent with the data. 
This is a consequence of the different sign of the exchanged amplitudes
in the
PP and VP multiplets \cite{lipkin,cr}, a feature also 
present in other analyses \cite{acgk,bn}.

\section{The scalar nonet in $B$ decays}

\subsection{Branching ratios: expectations and observations}

We now turn to the main part of our paper, the study of scalar particle
production in B-decays. This is motivated by
the remarkably strong signal observed for the scalar meson $f_0(980)$
by the BELLE  \cite{belle} (recent update \cite{belle3}) and
BaBar Collaborations \cite{babar1}
in the decay $B^+\to K^+\pi^+\pi^-$ 
where almost one half of the total rate above background falls into this
sub-channel
\begin{equation}
{\cal B} (B^+\to K^+f_0(980);\ f_0\to \pi\pi)\ \approx \
15 \times 10^{-6},
\label{f0krate}
\end{equation}
for more details, see Table \ref{tab:expscalars} below.
This large fraction for  $K^+f_0(980)$ is comparable to pseudoscalar decays
and 3 times larger than $K^+\rho^0$. Such a large rate 
could be taken as a first hint at the gluonic
affinity of this meson as well, but
it is clear that a more definitive answer requires an analysis 
similar to the one with $\eta'$ for scalar (S) particles
as well on the basis of their flavour classification.
Here we assume $f_0(980)$ to be a member of a $U(3)$ nonet.

\begin{table}[t]
\parbox[h]{15.3cm}{
\caption{Dominant contributions for 
$B$ decays into scalar (S) + pseudoscalar (P) or
vector (V) particles: penguin amplitudes $p_{XY}$ (normalized to 1
in each sector as in col. 3 of 
Table \protect\ref{tab:pseudosc}), exchange  and
gluonic amplitudes  $\beta_{PS},\beta_{VS}$ 
and $\gamma_{PS},\gamma_{SP},\gamma_{VS}$ 
resp. with scalar mixing angle $\varphi_S$; in brackets results for
 $\sin \varphi_S=1/\sqrt{3}\ (\varphi_S\sim\varphi_P)$;
cols. 3,6: upper sign for $B^0$,
lower sign $B^+$.}
$
\begin{array}{llcllc}
\hline\hline
B^0\to & B^+\to & \text{normalization to}&
  B^0\to & B^+\to & \text{normalization to}\\
P+S & P+S &  p_{PS} & V+S & V+S &  p_{VS} \\
 \hline
K^+a^- & K^0 a^+ & 1 & K^{*+}a^- & K^{*0} a^+ & 1 
\\
K^0a^0 & K^+ a^0 & \mp\frac{1}{\sqrt{2}} & K^{*0}a^0 & K^{*+} a^0 &
   \mp\frac{1}{\sqrt{2}} 
\\
K^0f_0 & K^+f_0 & \frac{1}{\sqrt{2}}(1+2\gamma_{PS})\sin \varphi_S &
  K^{*0}f_0 & K^{*+}f_0 & \frac{1}{\sqrt{2}}(1+2\gamma_{VS})\sin \varphi_S\\
       &        & \quad +(\beta_{PS} + \gamma_{PS})\cos \varphi_S &
       &        & \quad  +(\beta_{VS} + \gamma_{VS})\cos \varphi_S  \\
     & & \lbrack \frac{1}{\sqrt{6}}(1+2\beta_{PS}+4\gamma_{PS}) \rbrack &
     & & \lbrack \frac{1}{\sqrt{6}}(1+2\beta_{VS}+4\gamma_{VS})\rbrack \\
K^0f_0' & K^+f_0' & \frac{1}{\sqrt{2}}(1+2\gamma_{PS})\cos \varphi_S & 
     K^{*0}f_0' & K^{*+}f_0' & \frac{1}{\sqrt{2}} (1+2\gamma_{VS})\cos
      \varphi_S\\
      & & \quad   -(\beta_{PS} + \gamma_{PS})\sin \varphi_S &
      & & \quad  -(\beta_{VS} + \gamma_{VS})\sin \varphi_S  \\
     & &  \lbrack \frac{1}{\sqrt{3}}(1-\beta_{PS}+\gamma_{PS}) \rbrack &
        & & \lbrack \frac{1}{\sqrt{3}} ( 1-\beta_{VS}+\gamma_{VS})\rbrack \\
\pi^-K^{*+}_{sc} & \pi^+K^{*0}_{sc} & \beta_{PS} & 
    \rho^- K^{*+}_{sc} &  \rho^+ K^{*0}_{sc} & \beta_{VS}
\\
   \pi^0K^{*0}_{sc} & \pi^0K^{*+}_{sc} & \mp\frac{1}{\sqrt{2}} \beta_{PS} &
   \rho^0K^{*0}_{sc} & \rho^0K^{*+}_{sc} & \mp\frac{1}{\sqrt{2}} \beta_{VS}
\\
\eta K^{*0}_{sc} & \eta K^{*+}_{sc} &
\frac{1}{\sqrt{3}}(-1+\beta_{PS}+\gamma_{SP})&
     \omega K^{*0}_{sc} & \omega K^{*+}_{sc} & \frac{1}{\sqrt{2}} \beta_{VS} 
\\
\eta' K^{*0}_{sc} & \eta' K^{*+}_{sc} &
    \frac{1}{\sqrt{6}}(2+\beta_{PS}+4\gamma_{SP})&
   \phi  K^{*0}_{sc} & \phi K^{*+}_{sc} & 1\\
\hline\hline
  \end{array}
$
\label{tab:scalars}
}
\end{table}

In Table \ref{tab:scalars} we have written down the amplitudes for
the decays $B\to PS$ and $B\to VS$ in the approximation as above,
keeping only
the QCD penguin amplitudes ($p_{PS},p_{SP},p_{VS},p_{SV} $)
and gluonic amplitudes ($s_{PS},s_{SP},s_{VS}$) 
for the respective multiplets, see Table 1.
We denote the states of  the scalar nonet by $a_0,\ f_0,\ K^*_{sc},$ and 
$ f_0'$, where the isoscalar states are mixed from strange and non-strange
components as
\begin{equation}
f_0=n\bar n \sin \varphi_S + s\bar s \cos \varphi_S,\quad
f_0'=n\bar n \cos \varphi_S - s\bar s \sin \varphi_S
\label{mixings}
\end{equation}
with $n\bar n=(u\bar u +d\bar d)/\sqrt{2}$ and mixing angle
$\varphi_S$. 
For a given nonet of scalar states 
Table \ref{tab:scalars} predicts the corresponding 
pattern of decay rates in terms of these parameters. 

For the $PS$ decays the parameters are:
the normalisation of the penguin amplitude $p_{PS}$ 
and  $\gamma_{PS},\
\gamma_{SP},\ \beta_{PS}$ and for $VS$ decays the normalization $p_{VS}$ and
 $\gamma_{VS},\beta_{VS}$. According to our experience with 
the $PP$ sector we may 
assume in the beginning real parameters $\beta,\ \gamma$ and
an equal recombination probability for exchanged
processes and therefore restrict to $|\beta_{PS}|^2=1$.  Furthermore, we
assume, as in case of pseudoscalars, that the replacement of one
pseudoscalar by one vector meson keeps the parameters $\gamma, \beta'$
unaltered and therefore we choose
\begin{equation}
\beta_{PS}=\pm1, \quad \gamma_{VS}=\gamma_{PS}, 
   \quad \beta_{VS}=-\beta_{PS},   
\end{equation}
where the opposite sign for $\beta_{VS}$ comes from
the spin factor $(-1)^L$.

Next we ask whether
besides the very clear signal of $f_0(980)$ there is any 
evidence for production of other scalars in the data from
BELLE (Ref. \cite{belle} and the recent update, still preliminary, 
with higher statistics~\cite{belle3}) and BABAR~\cite{babar1}:
 
\begin{table}[ht]
\caption{
$B$ decays  into scalars measured by BELLE \protect\cite{belle3} 
 and BABAR \protect\cite{babar1} showing statistical and systematic as well
as model errors).}
$
\begin{array}{lll}
\hline
{\cal B} (B^+\to K^+f_0(980))
          &(10.3\pm1.1^{+1.0+0.2}_{-0.9-1.9})\times 10^{-6}
          & \protect\cite{belle3}\\
  \qquad \times{\cal B} ( f_0\to \pi^+\pi^-) 
          &(9.2\pm1.2\pm0.6^{+1.2}_{-1.9}\pm1.6)\times 10^{-6}
          & \protect \cite{babar1}\\
{\cal B}(B^+\to K^{*0}_0(1430)\pi^+)
          & (25.0\pm1.6^{+2.4+0.0}_{-2.1-1.5})\times 10^{-6}
          & \protect\cite{belle3}\ \text{Fit $C_0$/I}\\
  \qquad \times{\cal B} (K^{*0}_0\to K^+\pi^-)  
 & (6.00\pm0.84^{+0.58+0.33}_{-0.52-0.43})\times 10^{-6} 
         & \protect\cite{belle3}\ \text{Fit $C_0$/II} \\ 
   & (25.1\pm2.0\pm2.9^{+9.4}_{-0.5}\pm4.9)\times 10^{-6}
           & \protect\cite{babar1}\\
{\cal B} (B^+\to K^+f_0(1500))
   & (18.5\pm0.5)\times 10^{-6} & \protect\cite{belle3}\ \text{Fit $B_0$/I} \\
 \qquad\times {\cal B} (f_0\to K^+K^-)  & (1.3\pm0.2)\times 10^{-6} & \protect\cite{belle3}\ 
             \text{Fit $B_0$/II}\\ 
\hline
\end{array}
$
\label{tab:expscalars}
\end{table}

\noindent\underline{$K^{*}_0(1430)$} \\
A higher mass $K^*_0$ has been seen
by BELLE  and by BaBar 
 (BaBar quotes ``higher $K^*$'' which includes higher
spin states in this mass range) with rates reproduced in Table
\ref{tab:expscalars}.
Note the considerable difference in the two solutions by BELLE.

\noindent\underline{$f_0(1500)$ and broad ``background'' 
in $K\overline{K}$ and $\pi\pi$}\\
In the first
publications by BELLE \cite{belle,belle2} 
there was a broad low mass enhancement
in the  $K^+K^-$ and $K^0 \overline{K^0}$ mass spectra 
without any further structure. The 
recent preliminary results \cite{belle3}, show a qualitatively
different picture: there is a sharper peak at 1500 MeV
in the $K^+K^-$ mass spectrum, whose
mass and width agree with $f_0(1500)$, above a broad bump (``background'').
In the $\pi^+\pi^-$ mass spectrum a broad ``background'' can be seen as well,
however, there is no comparable peak at 1500 MeV
which at first makes the identification with
 $f_0(1500)$ difficult in view of the known branching ratio
${\cal B}(f_0(1500)\to K\overline K)/{\cal B}(f_0(1500)\to \pi\pi)=0.241\pm0.028$ \cite{pdg}   
based on the measurements \cite{obelix,cbc}. We will argue below that
the identification with $f_0(1500)$ and its branching ratio is possible if
the constructive and destructive interferences respectively
 with the ``background'' are taken into account.
In the $K^+K^-$ channel BELLE quotes for the resonance rate in 
two solutions $B_0/I$ and 
$B_0/II$ the quite different  
fractions 60.8\%  or 4.4\% of the charmless $K^+K^+K^-$ final state
from which we derive (with the errors from the fractions) the numbers in
Table \ref{tab:expscalars}.

\noindent\underline{$f_0(1370)$}\\
Belle \cite{belle3} also quotes the fractions of a small enhancement near
1300 MeV in $\pi^+\pi^-$ as $f_X(1300)$ which could be a signal from $f_0(1370)$. \\
\underline{$\kappa(850)$}\\
The existence of this state with parameters determined by the 
 E791 collaboration  
\cite{e791} (Mass 797 MeV, width 410 MeV) could not be verified
by BELLE \cite{belle3}. 
If a resonance is fitted it would have the much larger width of 2.27 GeV.
The best description of the decay $B^+\to K^+\pi^+\pi^-$ by BELLE includes a 
broad background in $K\pi$ with constant phase.\\
\underline{$\sigma(600)$}\\
One may ask also whether there is any evidence for the $\sigma(600)$
particle which is seen in some experiments as peak near the $\pi\pi$ 
threshold. 
The recent
BELLE data \cite{belle3} do not show any peak below the $\rho$ meson above
background whereas the statistics in BaBar is too low to make definite
statements. 
It would be interesting to have some limits or estimates of the
$B$ branching fractions into these hypothetical mesons $\sigma$ and $\kappa$
to be compared with the $f_0(980)$ branching fraction.\\
\underline{$a_0(980)$}\\
An important state in scalar spectroscopy is the $a_0(980)$ which directly
measures the penguin amplitude within the scheme of Table \ref{tab:scalars}.
A possible decay mode is  $a\to K\overline K$, which yields a peak
just above $K\overline K$ threshold.
It is interesting to note that the 
decay $B^0\to K^+K^- K^0$ studied by BELLE \cite{belle2} clearly shows
such a threshold enhancement near 1000 MeV in $K^+K^0$ and $K^-K^0$,
which could be due to $a_0(980)$ decay. 
However, there is a large background in this region from non-$B$ decays
and there is no claim for observation of $a_0(980)$ by the 
BELLE collaboration.\footnote{We thank A. Garmash for 
clarification of this point.} 
A determination through the decay
mode  $a_0(980)\to \eta\pi$ would be interesting, also for 
the heavier $a_0(1450)$.
Such studies would decide 
whether the nonet partner of $f_0(980)$ is $a_0(980)$ or $a_0(1450)$.\\

The above list suggests the observation of scalar states heavier than
$f_0(980)$ whereas there is no clear evidence yet for the observation of the
states of lower mass.
This is along route A which we discuss next.

\subsection{Comparison with heavy nonet scheme}
In this scheme $f_0(980)$ is the lightest member of the nonet.
Specifically, in our version \cite{mo} the members are as in (\ref{moscnonet})
and in addition there is a broad glueball.
The scheme \cite{klempt} is similar, with preference for $a_0(1450)$
but without glueball. In both schemes $f_0(980)$ and $f_0(1500)$ are mixed
as given by (\ref{mononet}).

We begin with a discussion of the production of $f_0(1500)$.
The peak in the $K^+K^-$ mass spectrum and a lack of a signal in $\pi^+\pi^-$ 
finds a natural interpretation in the quark structure  (\ref{mononet})
with the negative sign between nonstrange
and strange components \cite{mo,klempt} together with the glueball
interpretation of the background \cite{mo}. This negative relative 
sign transfers into the decay
amplitudes of $f_0\to K\overline K$ on one side and $f_0\to\pi\pi$ or 
$f_0\to\eta\eta$ on the other
side, whereas for the glueball the decay amplitudes have the same sign for all
pseudoscalar pairs. Therefore, if the interference $f_0(1500)$ - background
is constructive for $K\overline K$ we expect it to be destructive in
$\pi\pi$.
We recall that such a difference
has been observed already
in the channels $\pi\pi\to K\overline K$ 
and $\pi\pi \to \eta\eta$ in their interference with the broad background
according to our earlier analysis \cite{mo}.

In order to check the validity of this proposal quantitatively, for the
given decay ratio of $f_0(1500)$ into $K\overline K$ and $\pi\pi$
we compare the data with
a simple parametrisation of the decay amplitudes in terms of
$f_0(980)$, $f_0(1500)$ and the broad 
background which we interpret as glueball 
($gb$). The new feature of this fit in 
comparison to the original BELLE work \cite{belle3} is the common description
of both the $\pi\pi$ and $K\overline K$ channels and the inclusion of a 
background amplitude in both channels with a phase according to a 
broad resonance. 
A more complete analysis
should provide a fit of the two-dimensional 
Dalitz plot density taking into account the
interferences of resonances in the crossed channels. Here we are only
interested in the low mass region below 1700 MeV. In this region,
we may neglect the small contribution from crossed channel resonances
in $K^+\pi^+\pi^-$ ($K^*$); in $K^+K^-$ 
there seems to be some additional contribution below the
low mass enhancement 
which could come from charmed meson sources
in the crossed channel or some other background. 

In any case, we consider only  single channel 
amplitudes (with interactions in $\pi^+\pi^-$ or $K^+K^-$) 
which we describe as superposition of the three resonance contributions.
The spectrum in the pair mass $m$ is then given by:
\begin{gather}
\frac{d\Gamma}{dm} = |c_1|^2 q p
                |T_{gb}+c_2T_{f_0}S_{gb} + c_3T_{f_0'}S_{gb}|^2
          \label{amplitude}\\
T_a=\frac{m_a\Gamma_a}{m_{a}^2-m^2-im_{a}\Gamma_{a}(1+G_a(m))},\quad
a=gb,f_0,f_0'\\
S_{gb}=e^{2i\delta_{bg}};\qquad T_{gb}=| T_{gb}|e^{i\delta_{bg}}.
\end{gather}

The superposition of a narrow resonance with a broad background we describe
by the rotation of the resonance term in the complex plane 
with the background $S$ matrix,
$S=e^{2i\delta_{bg}}$, which is consistent with unitarity in the elastic
region as is well known; here we also apply it in the inelastic region.
In (\ref{amplitude})
$q$ is the hadron momentum in the $h^+h^-$ restframe and $p$ the momentum of
the $h^+h^-$ pair in the $B$ restframe.

The resonances included here cannot be given by a simple Breit Wigner form
in the considered mass range because of the distortion by inelastic
 $K\overline K$,
$\eta\eta$ and $4\pi$ ($\rho\rho)$ thresholds. In our present exploratory
study we do not attempt a full treatment of unitarity effects of the
coupled channel system, rather we allow for 
an energy dependent width which takes into account the $K\overline K$ threshold.
As we are unable to fix the shape function $G_a(m)$ in our fitting of
projected densities we choose, for definitness, a form which ties the
amplitude at energies $>1.6$ GeV back to the original resonance circle and is 
otherwise adjusted to give a reasonable representation 
of the mass spectra. For $f_0(980)$ and the glueball ($gb$) we chose
\begin{align}
K\pi\pi: \qquad
 & G_a(m)= \Theta(m-2m_K)\varepsilon_\pi x  \exp(-(m/m_1)^{11})\\
K\overline K K: \qquad
 & G_a(m)= \Theta(m-2m_K)\varepsilon_K x  \exp(-(m/m_2)^{11.5})
\end{align}
with $x = \sqrt{1-4m_K^2/m^2}$ and parameters 
$\varepsilon_\pi=0.9$, $\varepsilon_K=2.5$, $m_1$=1.28 GeV, $m_2$=1.37 GeV.
For
$f_0(1500)$, further away from the important thresholds, we put $G_a(m)=0$.
The resonance parameters and production amplitudes are taken as in Table
\ref{tab:resonances}. Concerning the glueball parameters we note that
the elastic $\pi\pi$ scattering amplitude in the
intermediate energy region around 1 GeV can be represented by a
superposition of T-matrix poles from $f_0(980)$ and a broad state
with mass near 1000 MeV and a width of 500-1000 MeV as is known since long
\cite{hyams,mp}. More recent analyses prefer a slightly higher mass in 
the region 1200 - 1600 MeV 
for this broad state by fitting to a larger number of channels \cite{anis}. 

\begin{table}[ht]
\caption{Parameters for resonances and their production amplitudes used in
the description of the  $\pi\pi$ and $K\overline K$ mass spectra in Eq.
(\protect\ref{amplitude}).
}
\begin{center}
$
\begin{array}{lcccc}
\hline\hline
\text{state} & \text{mass [GeV]} & \text{width [GeV]} & T(\pi^+\pi^-) &
                                                T(K^+K^-)\\
f_0(980)     & 0.99       & 0.05    &  c_2=5.00\ e^{0.1 i\pi}  
                    & c_2=2.75\ e^{0.1 i \pi}   \\
f_0(1500)    & 1.48      & 0.10    &  c_3=1.35\ e^{1.50 i \pi}  
                    & c_3=0.34\ e^{0.75 i \pi}  \\
gb           & 1.15       & 0.70     & 1    &  1  \\
\hline\hline
\end{array}
$
\end{center}
\label{tab:resonances}
\end{table}

\begin{figure}[b!]
\begin{center}
\vspace{-1.0cm}
\hspace*{-1.25cm}
\includegraphics[angle=0,width=18cm]{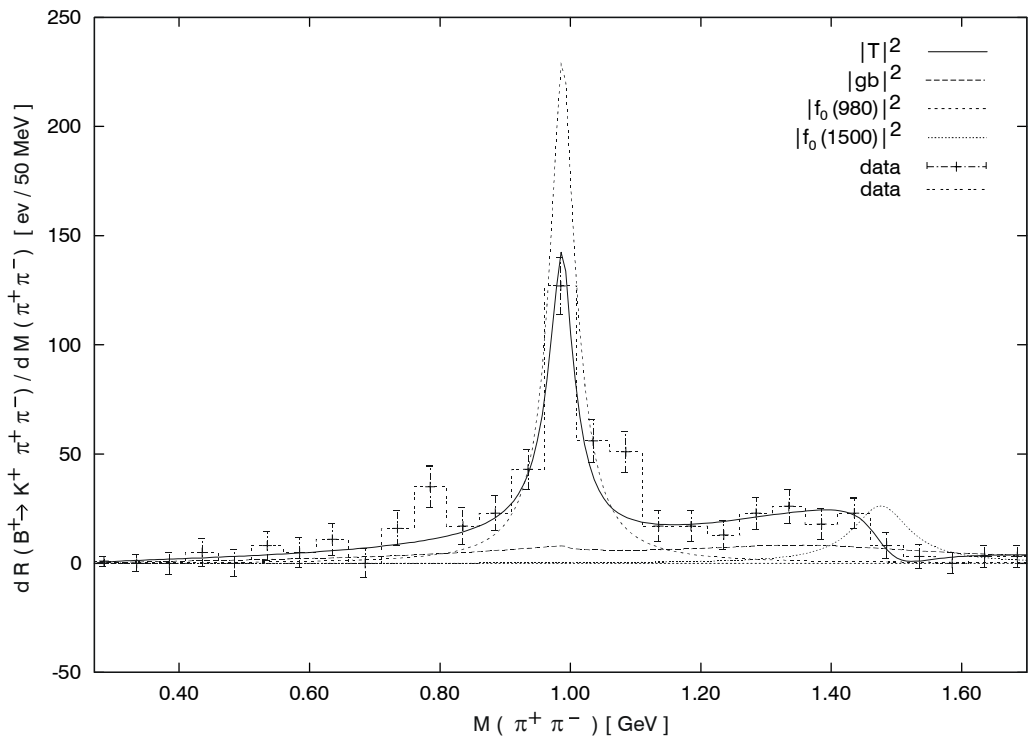} 
\end{center}  
\vspace{-1.0cm}
\caption{$\pi^+\pi^-$ mass spectrum in $B$-decays 
as measured by BELLE
\protect\cite{belle3} in comparison with a model amplitude $|T|^2$ 
of the coherent superposition of $f_0(980),\ f_0(1500)$ and a glueball ($gb$)
 forming the broad background. 
Also shown are the individual resonance terms
$|T_R|^2$. The background ($gb$) in this fit
interferes destructively with both $f_0(980)$
and $f_0(1500)$ consistent with the model 
(Eqs. (\protect\ref{mononet}) and (\protect\ref{amplitude})).
}
\label{fig:pipm}
\end{figure}   

\begin{figure}[ht!]
\begin{center}
\vspace{-1.0cm}
\hspace*{-1.275cm}
\includegraphics[angle=0,width=18cm]{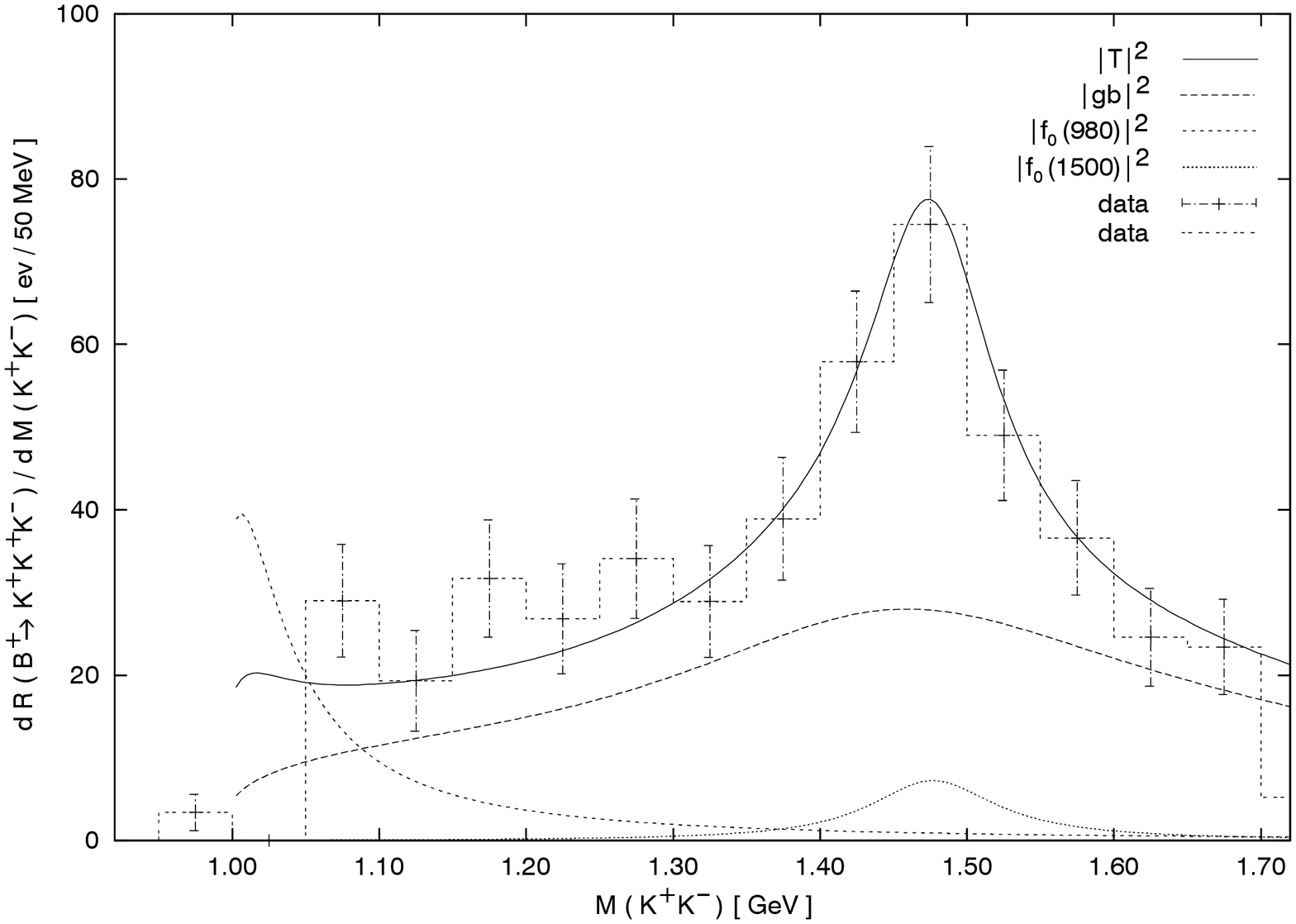} 
\end{center}  
\vspace{-1.0cm}
\caption{$K^+K^-$ mass spectrum in $B$-decays 
as measured by BELLE
\protect\cite{belle3} in comparison with the model amplitude, see Fig.
(\protect\ref{fig:pipm}). Here $f_0(1500)$ interferes constructively
with the background.
}
\label{fig:kpm}
\end{figure}   

The amplitudes were obtained by adjusting the
parameters ``by eye'' 
to investigate the result from superimposing the 3 resonances with
relative phases where we kept the branching ratio of $f_0(1500)$
at the value suggested by the PDG
(i.e. $K^+K^-/\pi^+\pi^- = 3/4 \times 0.241$) and to describe the data at 
least qualitatively. We do not include the production of the $\rho$ meson
which is clearly present \cite{belle3}.
The results with
these parameters are shown in 
Figs. \ref{fig:pipm},\ref{fig:kpm} in comparison with
the BELLE data.

We note that the relative phase between $f_0(1500)$ in $\pi\pi$ and $K\bar
K$ is $0.75\pi$, close to the expected value $\pi$ for a near octet
resonance, see Table \ref{tab:resonances}.
This phase difference causes the peak and dip in $K^+K^-$ and $\pi^+\pi^-$
respectively which we consider as an important further confirmation of the 
near octet nature of this meson which does not allow a sizable glueball
admixture (the glueball would contribute with the same sign to both particle
pairs, see also \cite{mo,momont2}). 

In our parametrisation the phases for
$f_0(980)\to \pi^+\pi^-$ and $f_0(980)\to K^+ K^-$ decays 
are about the same, but these phases 
are less well determined by the data. Note that the relative phase
between both $f_0(980)$ amplitudes is about zero relative to the background
in Table \ref{tab:resonances}.
The glueball  amplitude representing the broad background has phase 
$\delta_{gb}\approx\pi/2$ near the pole position around 1000 MeV. Therefore the 
interference with $f_0(980)$ is $destructive$ because of the phase
factor $S\approx e^{i\pi}$ in (\ref{amplitude}). This is well known for elastic $\pi\pi$
scattering where $f_0(980)$ produces a dip instead of a peak in the mass
spectrum.  In our present fit
the small background doesn't produce a dip but it
reduces the size of the $f_0(980)$ peak in
Fig. \ref{fig:pipm}; however, the phase is not so well determined and a
constructive interference cannot be excluded. This requires a study of the
lineshape with higher accuracy. Such a
 phenomenon is met also in the decay $J/\psi\to \phi\pi\pi$
\cite{dm2} where the interference of $f_0(980)$ with the background is
destructive. This is derived from the asymmetric shape of $f_0(980)$.

Our description also reproduces approximately the small enhancement
near 1300 MeV which is generated by the interference of all three states.
On the other hand, it
has been interpreted by BELLE as due to $f_0(1370)$. 
Whether $f_0(1370)$
really exists as a particle 
is still controversial. The PDG does not confirm any branching
ratio. If the measurement of the decay ratio
${\cal B}(f_0(1370)\to K\overline K)/{\cal B}(f_0(1370)\to\pi\pi)=
0.46\pm 0.15\pm0.11$ \cite{omeg} 
is taken for granted,
one should see in $B$ decays in both channels 
an effect of size comparable to $f_0(1500)$. The $K^+K^-$ spectrum does not
provide immediate evidence for such a situation.

\begin{table}[t] 
\caption{Decay branching ratios for resonant contributions in $B^+\to K^+\pi^+\pi^-$ and
$B^+\to K^+K^+K^-$ for masses $m<1.72$ GeV from our analysis using the branching ratio equivalents
of $0.05417\times 10^{-6}$ and $0.03657\times 10^{-6}$ per event 
(as concluded from \cite{belle3}) for both
final states respectively.
}
\begin{center}
$
\begin{array}{lccc}
\hline\hline
\text{channel} &  gb\ [10^{-6}]  & f_0(980)\ [10^{-6}]  & f_0(1500) \
    [10^{-6}] \\
\hline
\pi^+\pi^-     & 7.4   & 19.3    & 4.3\\
\pi\pi         & 11.1 & 29.0   & 6.5 \\
K^+K^-         & 10.1 & 2.1 & 0.70  \\
K \overline K       & 20.2 & 4.2 & 1.4   \\
\text{all}     &   -   & 33.2 & 20.3 \\
\hline\hline
\end{array}
$
\end{center}
\label{tab:b-ratio}
\end{table}

Finally, we determine the decay rates for the resonances studied and compare 
with our expectations from Table \ref{tab:scalars}. From our analysis we
obtain the number of events for each state (the area under the
curves for our three states $|T_a|^2$ in Figs.
\ref{fig:pipm} and \ref{fig:kpm}). Relating the total event numbers in the two
channels 
with the corresponding branching ratios for charmless decays 
we arrive at the numbers in Table
\ref{tab:b-ratio}. In order to obtain the total rates we added
for $f_0(980)$ the observed decays in $\pi\pi$ and $K\overline K$. 
In case of $f_0(1500)$ a complete sequence
of branching ratios is not available from the PDG. If we use the ratios
determined by the PDG \cite{pdg} and estimate the ratio 
${\cal B}(f_0\to 4\pi)/{\cal B}(f_0\to 2\pi)\approx 1.7$ from their listing
we obtain
the following estimate for the branching ratios
\[
f_0(1500):\ 4\pi\ (54.2\%),\ 2\pi\ (32.0\%),\ K\overline K\ (7.8\%),
\ \eta\eta\ (4.2\%),\  \eta\eta'\ ( 1.8\%). \label{rates1500}
\]
From the $2\pi$ decay we obtain the branching ratios 
of Table \ref{tab:b-ratio}. 

We note that our
result for $f_0(980)\to \pi^+\pi^-$ is about twice as large as in
Table \ref{tab:expscalars}  
whereas our result for $f_0(1500)\to K^+ K^-$ is smaller by the same
amount compared to Fit $B_0/II$ 
and much smaller than Fit $B_0/I$ in Table \ref{tab:expscalars}. 
Note that in case of $f_0(1500)$ 
we fit the
$\pi\pi$ spectrum as well 
with the $K\overline K/\pi\pi$ ratio kept fixed (at PDG value) 
and the large rate of Fit $B_0/I$ seems to be excluded.

If we had taken the fit results by BELLE on $f_0(980)$ and $f_0(1500)$ (Fit
$B_0/II$) instead of ours in  
in Table \ref{tab:b-ratio} we would have obtained the corresponding
rates  $17\times 10^{-6}$ and  $33\times
10^{-6}$ (instead of  $33.2\times 10^{-6}$ and $20.3\times 10^{-6}$).
The different results may come partly from a different 
treatment of phases, especially our moving background phase; we also remark
that
our analysis is using an approximation by working only with projections.
We estimate that the present numbers for the rates have a
model uncertainty by about a factor of two.

Finally, we compare our results with the expectations from 
Table \ref{tab:scalars} based on symmetry relations for the nonet.
We derive the predictions using the mixing angle sin $\varphi=1/\sqrt{3}$
as in our model \cite{mo} and in \cite{klempt}.
We are using the simplifications motivated by the experience with
pseudoscalars as outlined in the last subsection. 
We assume real $\gamma_{PS}$ and restrict to $\beta_{PS}=\pm1$.
Then we can use the rates for $f_0\equiv f_0(980)$ and 
$f_0'\equiv f_0(1500)$ to determine the 
two parameters $|p_{PS}|^2$ and $\gamma_{PS}$. The result for $|p_{PS}|^2$
may then be compared
with the measurement of the $K^*_0(1430)\pi$ rate. We allow for both signs of
$\beta_{PS}$ and obtain for each case two solutions from the predictions
for the rates following from Table \ref{tab:scalars} (amplitudes in brackets)
and these results are presented in Table~\ref{tab:scamp}.

\begin{table}[th]
\caption{Parameters $|p_{PS}|^2$ (in units of $10^{-6}$) and  $\gamma_{PS}$ 
in the decay of $B$ into scalar particles using the
symmetry relations in Table \protect\ref{tab:scalars} and the total 
branching ratios for $a_0,\ f_0\equiv f_0(980)$ and $f_0'\equiv f_0(1500)$ 
from Table 
\protect\ref{tab:b-ratio}. There are two solutions for each value of
$\beta_{PS}$; further predictions of branching ratios 
 (in units of $10^{-6}$) 
using for $K^*(890)$: $\gamma_{VS}=\gamma_{PS}$ and $\beta_{VS}=-\beta_{PS}$
and $p_{VS}/p_{PS}=0.437$; Sol. 2 excluded, see text.
}
\begin{center}
$
\begin{array}{lccccccc}
\hline\hline
\text{Sol.} &  \beta_{PS}  &|p_{PS}|^2 & \gamma_{PS}\quad & K^0a^+_0 & 
         K^{*+}f_0 &   K^{*+}f_0' & K^{*0}a_0^+ \\
\hline
1           &   +1   & 32.0   & -1.38 &32.0 &13.2 & 1.8 & 14.0\\
2           &   +1   & 234.   & -0.51  &- &- &- &-\\
3           &   -1   & 11.0   & 1.33   &11.0 &55.4 &2.8  & 4.8\\
4           &   -1   & 42.1   & -0.30   &42.1 &9.9 &0.6 & 18.4\\
\hline\hline
\end{array}
$
\end{center}
\label{tab:scamp}
\end{table}
With $|\beta_{PS}|=1$ we can compare these results for $|p_{PS}|^2$ directly 
to the rate for $B^+\to K^{*0}_0(1430)\pi^+$ according to Table \ref{tab:scalars}.
Correcting the numbers in Table \ref{tab:expscalars}  for neutral decays
one finds for $K^{*0}_0\pi^+$ the rates about $38\times 10^{-6}$ or 
9$\times 10^{-6}$. 
Then we can exclude our Sol. 2 in Table~\ref{tab:scamp} whereas
Sol. 1 and 4 are near Fit $C_0/I$ in Table \ref{tab:expscalars} and Sol. 3
near Fit $C_0/II$. A more detailed investigation of the Dalitz plot
densities  
could possibly distinguish between Solutions $C_0/I$ and $C_0/II$.
Another test is possible through measurement of the $K^0a^+_0$ decay
which is given directly by $|p_{PS}|^2$.

We note that Sol. 4 in Table~\ref{tab:scamp} has similar parameters
to the pseudoscalar amplitudes
$\beta_{PP}=1$ and $\gamma_{PP}=0.44$, except for their opposite sign. 
On the other hand, Sol. 1 and 3 have a three times larger gluonic amplitude
which would not be theoretically expected. If we had taken the BELLE Fit results instead
of ours in Table \ref{tab:b-ratio} we had obtained
the corresponding numbers to Sol. 4 as $|p_{PS}|^2=30.5\times 10^{-6}$
and $\gamma_{PS}=-0.20$. 

In any case, the results obtained are not 
in contradiction with the suggested scalar
nonet of heavier particles along Route A. Actually, our specific approach
\cite{mo} with nonet (\ref{moscnonet}),(\ref{mononet}) and broad glueball 
explains naturally the 
new observations from $B$ decays. Further results concern the predictions 
for production of $a_0$
and of scalars together with vector mesons, especially $f_0,\ f_0'$. 
These predictions are shown also
in Table~\ref{tab:scamp} for the 3 remaining solutions. Again we make the
simplified choices motivated by the experience with the pseudoscalars,
as outlined in the previous subsection, and assume
  $\gamma_{VS}=\gamma_{PS}$ and  $\beta_{VS}=-\beta_{PS}$ (the opposite sign 
for $\beta_{VS}$ comes from the spin factor $(-1)^L$), also we take
$p_{VS}/p_{PS}=0.661$. Then we expect a much
smaller rate for $K^* f_0(1500)$ than for $K^* f_0(980)$. The rate
$K^{*0}a^+_0$ determines directly the normalization $|p_{VS}|^2$.
These measurements will provide  further important tests of our picture.

\subsection{Comparison with light nonet scheme }
In this approach $f_0(980)$ falls into the same nonet with
$a_0(980)$ and also with $\sigma(600)$ and
$\kappa(850)$ whose existence are still under debate. If this nonet is built from $q\bar q$ states 
with mixing as in (\ref{mixings}) we can
apply the same discussion as before and explore the symmetry relations
of Table \ref{tab:scalars}. In case of a $qq\bar q\bar q$ model additional
degrees of freedom may come in which we do not consider here. 
As emphazised in the beginning of this
section, there are not yet any definitive
observations of these light scalars in $B$ decays, as they exist for
the heavier scalars, nor are there any limits for branching ratios.
Therefore we only indicate some possible scenarios to stress the potential 
of $B$-decay studies also for these light scalars.

In this case only the rate for $B^+\to K^+f_0(980)$ 
is available which we take from
(\ref{f0krate}). A natural choice is the mixing angle $\varphi_S=0$
which, according to (\ref{mixings}), corresponds to 
$f_0(980)$ being a pure  strange and $\sigma$ a pure non-strange
state. The decay amplitudes follow from Table \ref{tab:scalars}
and are listed for a few decay channels in Table \ref{tab:sigma}.

The simplest choice is to assume a production without gluonic
processes, i.e. $\gamma_{PS}=0$ (Scenario I)
We assume again
$|\beta_{PS}|=1$, as in the previous
discussions of pseudoscalar and scalar sectors. Then, with the $f_0$ rate as
input we can predict the other states of the multiplet. One finds, that
in this case the $K\sigma$ rate is 
1/2 of the $Kf_0$ rate and equal to the $K\kappa$
rate. This looks rather large for $\sigma$ in view of first 
results \cite{belle3}. 

We therefore also consider another Scenario II where we
introduce a gluonic coupling of $f_0$ such as to cancel the $K\sigma$ decay.
In this case the sign of $\beta_{PS}$ matters and we obtain 
for $K\kappa$ either $60\times 10^{-6}$ or  $6.7\times 10^{-6}$
where the first choice can presumably be excluded from the data
\cite{belle3}.
Furthermore, we present some 
predictions for decays with vector meson $K^*$. Again, the measurement of
the $a_0$ rate would be very useful as it fixes the normalizations for the
$PS$ and $VS$ multiplets.
We conclude that a measurement of the rates for $\sigma$ and $\kappa$ 
will be important for the discussion of their existence, their 
classification and their production mechanism.

\begin{table}[t]
\caption{Rates for some channels with scalars
 assuming $f_0(980)$ in a nonet with
$\sigma(600)$ and $\kappa(850)$ (mixing angle $\varphi_S=0$) and amplitudes
(in units of the penguin amplitudes $p_{PS},\ p_{VS}$)
for two scenarios with different gluonic component
$\gamma_{PS}$, using 
$\beta_{PS}=\pm1$. For $K^*$-decays we take $\beta_{VS}=-\beta_{PS}$, 
 $\gamma_{VS}=\gamma_{PS}$ and $p_{PS}/p_{VS}=0.661$. The $K^+\kappa^0$ 
and $K^0a_0^+$ rates
equal $|p_{PS}|^2$.
}
\begin{center}
$
\begin{array}{lcccccc}
\hline\hline
B^+\to  &  K^+f_0 & K^+\sigma & K^0a^+_0 & \pi^+\kappa^0 & K^{*+}f_0 & K^{*+}\sigma \\
\text{amplitude} & \beta_{PS}+\gamma_{PS}
    &\frac{1}{\sqrt{2}}(1+2\gamma_{PS}) & 1 & \beta_{PS} &
          \beta_{VS}+\gamma_{VS} & \frac{1}{\sqrt{2}}(1+2\gamma_{VS})\\
\hline
\text{Scenario I}: &  \gamma_{PS}=0 &  & & & \\
\text{rate [$10^{-6}$]} & 15.  & 7.5  & 15. &15. & 6.5  & 3.2\\
\hline
\text{Scenario II}: &  \gamma_{PS}=-0.5 &\beta_{PS}=+1 & &&\\
\text{rate [$10^{-6}$]} & 15. & 0. & 60. & 60. & 59.0 & 0. \\
              &   & \beta_{PS}=-1 & && \\
\text{rate [$10^{-6}$]} & 15. & 0. & 6.7 & 6.7 & 0.7& 0. \\
\hline\hline
\end{array}
$
\end{center}
\label{tab:sigma}
\end{table}

\subsection{Total rate for gluonic decays} 
Next we compare the gluonic production rates for $f_0,\ f_0'$ and 
$\eta,\ \eta'$ with the
total rate $b\to sg$ in (\ref{btosg}).
 CLEO \cite{cleo1} has measured the inclusive non-charm decay
${\cal B}(B\to \eta'+X)\ = \ (6.2^{+2.1}_{-2.6})\times 10^{-4}$,
where the signal refers to the region $2.0<p_{\eta'}< 2.7$ GeV
of the $\eta'$ momentum. Identifying the non-charm rate with $X_s$ according
to the SM and adding the exclusive $\eta'K$ rate we obtain
the inclusive rate 
${\cal B}(B\to \eta'+X_s)\ \sim \ 7.0\times 10^{-4}$,
so the ratio of the total inclusive $\eta'X_s$ over the exclusive $\eta'K$ 
rate is  $R_{\eta'}(\text{incl/excl})\approx 9$.
We take the gluonic part $3|\gamma_{PS}p_{PS}|^2$ as in (\ref{etaprgluon}) 
whereas for scalars we find from Table \ref{tab:scamp} the corresponding
result in the range $(11\ldots182)\times 10^{-6}$ with preference for the
lowest value.
 Then we find for the fully inclusive contribution of these  decays after
multiplication with $R_{\eta'}(\text{incl/excl})$
\begin{equation}
{\cal B} (B\to \eta,\eta',f_0,f_0')|_{\rm gluonic}\sim (0.2\ldots 2)\times
   10^{-4}
\label{glumeson}
\end{equation}
with preference for the lower value.  
Hence, these decays cannot contribute more than a small fraction
of the  expected $b\to s g $ rate of $5\times 10^{-3}$ 
\cite{greub}.
We will argue next that glueball production does provide the dominant
part of the $b\to sg$ decay originating from a genuine hard gluon composing
the associated local dimension 5 operator.

\section{Glueball production in $B$ decays}
Besides the observation of the strong $f_0(980)$ signal 
there is another interesting clear feature in the $B$ decays: the 
presence of a broad low mass $\pi\pi$ and $K\overline K$ 
enhancement in $B^+\to K^+\pi^+\pi^-$
and $B^+\to K^+ K^- K^+$ with spin $J=0$, 
observed by the BELLE collaboration \cite{belle,belle2,belle3}.

An important signal for the sizable S wave background is the interference
with known resonances. The appearence of the $f_0(1500)$ peak in $K^+K^-$
on one hand and its disappearence in $\pi^+\pi^-$ on the other hand
can be naturally explained by its constructive and destructive interference
with this background as discussed in the previous section. 

This interference is quite pronounced, as in elastic $\pi\pi$ scattering,
   where in contrast $f_{0}(980)$ and
$f_0(1500)$ appear both as dips in the broad background which we interpreted
in our earlier study \cite{mo} as destructive interference with the
broad glueball (``red dragon'').  
In the low energy
region below 1 GeV we assume that the $\pi\pi$ amplitudes are moving
as in elastic $\pi\pi$ scattering, at higher masses the movement of the phase is more
difficult to predict because of the inelastic channels. 
The study of interferences in the Dalitz plot, especially the crossing
regions of the known resonances - charmed and
uncharmed - with the background and among themselves 
could eventually show whether 
the movement of the background phases in  $\pi\pi$
and $K\overline K$ channels is consistent with an inelastic broad resonance.
An open question here is also the relevance of $f_0(1370)$ in these channels.
We argued in Ref. \cite{mo} that the Breit-Wigner phase motion 
     of an assumed resonance with moderate width at this energy has not
     been demonstrated clearly enough to require a definite resonance.
     In a recent analysis of the $f_0(1370)\to 4\pi$ channel 
 \cite{reinn} (as reported in \cite{klemptrev2}) it was found that in 
the considered mass range the phase motion was lower than expected 
from the Breit Wigner formula so that these data do not support the 
interpretation of $f_0(1370)$ as "normal resonance".  

Besides the prediction of a moving phase the glueball hypothesis has
also consequences for the decay fractions into different particles.
In order to relate different channels and to obtain an estimate of the total
glueball production rate we consider the following decay scheme. The
glueball decays first into the various $q\bar q$ pairs 
with equal amplitude
(possibly also into a pair of gluons)
\begin{equation}
gb\to u\bar u + d\bar d + s\bar s\quad (+gg). 
\label{gbdecay}
\end{equation}
The $gg$ pair could hadronize into two secondary glueballs 
after creation of another gluon pair. 
 Our $0^{++}$ glueball at energies above 1 GeV 
could decay as $gb\to \sigma\sigma$ where $\sigma(600)$ is
considered as the low mass part of the same $0^{++}$ glueball.
Then the main decay process would be $gb\to 4\pi$.
In the following we do not consider these decays further here. 
Each of the $q\bar q$ 
pairs in (\ref{gbdecay}) recombines with a newly created pair
$u\bar u$, $d\bar d$ or $s\bar s$ where $s\bar s$ is
produced with amplitude $S$ ($|S|\leq 1$). In this way the 2-body channels 
$gb\to q\bar q'+\bar q q'$ are
opened, at low energies just pairs of pseudoscalars. They are produced
with probabilities
\begin{equation}
\vspace{-0.8cm}
\begin{array}{cccccc} 
\pi^+\pi^- & \pi^0\pi^0 & K^+ K^- & {K^0 \overline K}^0 & \eta\eta &
     \eta'\eta'\\
 2 & 1 & 2 & 2 & 1 & 1 \\
 2& 1 & \frac{1}{2}|1+S|^2 & \frac{1}{2}|1+S|^2 & \frac{1}{9}|2+S|^2 &
  \frac{1}{9}|1+2S|^2\\
\label{gb2body}
\end{array}
\end{equation}

\vspace{-0.6cm}
\noindent The first row corresponds to $U(3)$ symmetry ($S=1$), the second
row to arbitrary $S$ (for numerical estimates we take $S=0.8$); 
 $\eta,\eta'$ mixing is assumed as
above. With increasing glueball mass 
the $q\bar q$ pairs can
decay also into pairs of vector mesons or of other states. The total
rates in (\ref{gb2body}) are assumed to remain unaltered but in general
``$\pi\pi$'' is meant to include $\rho\rho$ as well above the respective
threshold of about 1300 MeV.

We consider first the mass region 1.0-1.7 GeV.
In this region the pseudoscalars alone saturate the ``$K\overline K$'' 
rate in (\ref{gb2body})
as $K^*\overline K$ is forbidden by parity and $K^*{\overline K}^*$ is 
kinematically suppressed.
Another possible decay is $\eta\eta$, contributions from higher mass
isoscalars ($\omega\omega$) are only possible at the upper edge of the
considered mass interval. The decay  $\eta'\eta'$ is kinematically
forbidden.

In the region 1.1 - 1.3 GeV, away from the narrow resonances 
and from the major inelastic thresholds, we may compare 
the rates into the $\pi^+\pi^-$ and $K^+K^-$ channels
 with the theoretical expectations (\ref{gb2body}). 
From  Figs. \ref{fig:pipm} and \ref{fig:kpm} we obtain
$dR/dM(\pi^+\pi^-)\sim 20$ events/50 MeV in the mean  and 
$dR/dM(K^+K^-)\sim 30$ events/50 MeV
corresponding to $dR/dM \sim 1\times 10^{-6}$/50 MeV 
in both cases using the
${\cal B}$-equivalents in Table \ref{tab:b-ratio}. The glueball fractions in
our fits appear at the level of
about 1/2 of the experimental data 
in the mean in both cases in this mass range.
 Therefore the glueball decay rates in $\pi^+\pi^-$
and $K^+K^-$ are approximately equal, within about 20\%, 
consistent with the expectations $2:1.6$ (for $S=0.8$) from (\ref{gb2body}).

Above 1600 MeV the two spectra look quite different: whereas the $K\overline K$
spectrum decreases slowly, the $\pi\pi$ spectrum stays at a low level.
This may be explained by assuming the dominant decay of ``$\pi\pi$'' in
(\ref{gb2body}) proceeds into $4\pi$ (e.g. $\rho\rho$) states
in this mass range.

Next we estimate the total glueball rate. 
Here we start from our fit result for $gb\to K\overline K$ 
in Table \ref{tab:b-ratio}
refering to the mass interval $1.0\ldots1.72$ GeV. Including
the contribution from higher masses of about 55\% we obtain
${\cal B}(B^+\to K^+gb;\ gb\to K\overline K) \approx 31 \times 10^{-6}$.
This yields an estimate of the lower limit
for exclusive and inclusive 
branching ratios (correcting for other decay modes without $gb\ gb$ 
by a factor 2.2 from 
 (\ref{gb2body}) using $S=0.8$ by
neglecting  $\pi\pi$ decays below 1 GeV and $\eta\eta'$)
\begin{align}
{\cal B}(B^+\to gb(0^{++})+K^+) \gsim & \ 70 \times 10^{-6} \\
{\cal B}(B^+\to gb(0^{++})+X_s)\  \gsim & \ 0.6 \times 10^{-3},
\end{align}
here we used again the factor 9 from $K\eta'$ 
to estimate the fully inclusive rate.
 
By adding the gluonic pseudoscalar and scalar meson contributions
from (\ref{glumeson})
we estimate the lower limit (neglecting again  
the $gg$ decay mode in (\ref{gbdecay}) and
choosing Sol. 4 in Table \ref{tab:scamp}) 
for the total production
of observed gluonic scalar and pseudoscalar mesons as
\begin{equation}
{\cal B} (B^+\to gb(0^{++})+f_0+f_0'+\eta+\eta'+X_s)|_{{\rm gluonic}}
   \gsim 0.8\times  10^{-3}. 
\label{glumestot}
\end{equation}
This lower limit for gluonic production amounts to 
about 1/2 of the leading order result for the process
$b\to sg$ in (\ref{btosg}) and about 1/6 of the full rate obtained in NLO,
so it represents already a sizable fraction of the theoretically derived
value.
This result supports the expectation
that the $b\to sg$ rate will be saturated if
a few similar
processes with other $J^{PC}$ quantum numbers, especially the $0^{-+}$
and $2^{++}$ glueballs are included.

\section{Conclusions}

The main motivation for our 
investigation is the search for the lightest glueball
which is expected with $J^{PC}=0^{++}$ quantum numbers. To this end it is important
to obtain a full understanding of the light scalar sector, i.e. to establish
the lightest scalar nonet and its intrinsic mixing. An important role is played
by $f_0(980)$ which in different classification schemes is either the
lightest or the heaviest particle in this nonet (other options are being
discussed as well). 

In this paper we investigate the potential of 
$B$-decays which have the partial decay mode
$b\to sg$ in the search of gluonic objects.
This interest has been triggered by the undisputable observation of 
 $B\to K f_0(980) $ with a large rate comparable to the one for 
pseudoscalar particles.
This could imply a large gluonic affinity of this meson or a flavour singlet
nature similar to the~$\eta'$.

For the further investigation of $B$-decays we suggest a 
simple approximate scheme for 2-body decay rates 
based on the dominance of penguin
amplitudes with  an additional gluonic component; for each pair of
multiplets there are 3 amplitudes $p,\ \gamma p,\ \beta p$, 
the present analysis allows for
real $\gamma$ and $\beta=\pm 1$. 
 This model has been tested
first in the sector of decays  $B\to PP$ and $B\to VP$ 
where it corresponds to a simplified version of the previous approach
 \cite{cr}.  
Because of the (approximate) flavour symmetry of all $q\bar q$ 
in the decays $b\to s q\bar q$
all members of the nonet have a common component in the amplitude modified
by $\beta ,\gamma$ amplitudes. Exploiting this fact by comparing the rates
with the expectations in Table \ref{tab:scalars} should ultimately 
disclose the identity of the members of the lightest nonet associated with
$f_0(980)$.

The choice of the nonet with members $a_0(980), f_0(980),
K_0^*(1430), f_0(1500)$ with $f_0(980)$ as lightest particle
as in \cite{mo} and similarly in \cite{klempt} (with $a_0(1450)$ preferred)
can reproduce the observed phenomena concerning also 
other scalar particles not yet as
well established as $f_0(980)$. In particular, there is the remarkable
phenomenon of the $f_0(1500)$ signal in $K\overline K$, but apparently absent in
$\pi\pi$ despite the four times larger branching ratio. In our analysis this is
explained by the constructive and destructive interference respectively 
with the broad glueball (``background'', ``red dragon''). Such a behaviour is
expected from the near octet flavour composition of $f_0(1500)$. The
negative sign between the $s\bar s$ and $u\bar u+d\bar d$ component of
$f_0(1500)$ has been found before in inelastic $\pi\pi$ scattering \cite{mo}.
 If confirmed in the final analysis of the data 
it would seriously restrict the
possible glueball admixtures of $f_0(1500)$ which yield 
contributions of equal sign
to $\pi\pi$ and $K\overline K$ decay amplitudes.  

An important role in this classification is played by
$a_0(980)$ as the 
lightest isovector particle which directly determines the penguin
amplitude $p$ and therefore the overall normalization within one multiplet.
This particle can be identified from $B\to K\overline K K$ and
$B\to K\eta\pi$  decays.
Knowing this decay rate the other parameters in the model for the given decay
multiplets can be determined more directly. 
Alternatively, $a_0(1450)$ could be the nonet partner of $f_0(980)$.

The $a_0(980)$ rate would help in particular in a judgement about the classification with
$f_0(980)$ as heaviest particle in the nonet together with 
$\sigma$ and $\kappa$ in $B$-decays. At present there is no strong indication for 
the presence of these particles but a more dedicated analysis in the
determination of the respective decay rates is necessary.

The study of charmless $B$ decays into the various 
members of the nonet represents a
systematic approach to scalar spectroscopy which can be followed in great
analogy to the successful phenomenology of the decays into pseudoscalars.
A particularly interesting test is the comparison of decays $PS$ and $VS$ 
with $K$ and $K^*$
which are expected in some cases 
with quite different rates because of different signs of the parameter
$\beta$. Our first attempts indicate a different sign of the amplitudes
$\gamma,\beta$ in final states with scalar and pseudoscalar particles.
 
Finally, we come back to the question of gluonic meson production. First
there is the gluonic component in the production of isoscalar mesons
which is obtained for $\eta,\eta'$ already in previous analyses. We find such
components also in the scalar sector for $f_0(980)$. A more definitive
analysis can be done given the $a_0$ rates.

The presence of a coherent background can be derived from its  
interference with  $f_0(1500)$. In our glueball interpretation
we expect a production phase moving slowly with
energy according to a Breit Wigner amplitude with some modification
 by inelastic effects. 
It will be interesting to determine more accurately the interference between 
the background ($gb$) and $f_0(980)$ whether it is destructive (as in
$J/\psi\to \phi\pi\pi$ and in our model) or constructive; this has a big
influence on the derivation of the important $f_0(980)$ decay rate.
The $\pi\pi/K\overline K$ branching ratios of the background from our estimate 
are consistent with this
glueball hypothesis. Furthermore we predict a sizable $\eta\eta$ and, at
higher masses, a $4\pi$ decay rate. 

The broad background has also shown up in other gluonic processes 
like double pomeron production and $p\bar p$ annihilation, 
also with suppressed rate in $\gamma\gamma$ collisions
\cite{bp,mofrascati};
on the other hand, it has been difficult to see its sign in 
$J/\psi\to \pi\pi\gamma,
\ K\overline K\gamma$ (see discussion in \cite{mo}). 
Recent high statistics results by BES \cite{bes} 
on $K\overline K\gamma$, however, 
require a broad $0^{++}$ coherent background for a
good description of the data as well. We have no definitive 
explanation why
${\cal B}(J/\psi\to f_0(980)\gamma)$ is apparently much smaller than
${\cal B}(J/\psi\to \eta'\gamma)$ despite the similar 
quark structure proposed for $f_0(980)$ and $\eta'$. 
A possible explanation is a destructive interference with the background, 
seen in other processes as discussed, such that
the appearence of $f_0(980)$ is minimized, 
also a special effect in the
exclusive decay could be thought of. Therefore, it would be interesting
to see whether the similarity of $\eta'$ and $f_0(980)$ is recovered in
the fragmentation region of gluon jets as discussed in \cite{moringberg}.

It remains an interesting question how the $b\to sg$ decay is realized
by hadronic final states. The large rate for the  $0^{++}$ glueball we
obtain within our approach
suggests the intriguing possibility 
that it could be saturated by gluonic mesons. In the next 
step it will be interesting to search for the $0^{-+}$ glueball which
could decay into $\eta\pi\pi$ and $K\overline K\pi$.

\end{document}